\documentclass[prb,twocolumn,showpacs,superscriptaddress]{revtex4}
\pdfoutput=1 
\usepackage{epsfig}
\usepackage{amsmath,amssymb}
\usepackage{mathtools}
\usepackage{graphicx}
\usepackage{footnote}

\usepackage[normalem]{ulem}
\usepackage{color}

\newcommand{\ph}{\ensuremath{\mathrm{ph}}}
\newcommand{\pp}{\ensuremath{\mathrm{pp}}}
\newcommand{\phbar}{\ensuremath{\overline{\mathrm{ph}}}}

\newcommand{\ee}{\ensuremath{\mathrm{e}}}
\newcommand{\dee}{\ensuremath{{\hat d}^{\vphantom\dagger}}}
\newcommand{\ddgr}{\ensuremath{{\hat d}^{\mkern+1mu\dagger}}}
\newcommand{\dprime}{\ensuremath{\prime\mkern-1mu\prime}}

\usepackage[all]{xy}

\begin{document}

\ifcsname DIFaddbegin\endcsname%
  \newcommand{\sticky}[2] {\DIFaddend\textcolor{green}{{[}{\bf #1}: #2{]}}\DIFaddbegin}
\else
  \newcommand{\sticky}[2] {}
\fi

\title{Worm Improved Estimators in Continuous-time Quantum Monte Carlo}

\author{P.~Gunacker}
\affiliation{\small\em Institute for Solid State Physics, TU Wien, 1040 Vienna, Austria}
\author{M.~Wallerberger}
\affiliation{\small\em Institute for Solid State Physics, TU Wien, 1040 Vienna, Austria}
\author{T.~Ribic}
\affiliation{\small\em Institute for Solid State Physics, TU Wien, 1040 Vienna, Austria}
\author{A.~Hausoel}
\affiliation{\small\em Institute for Theoretical Physics and Astrophysics, University of W\"urzburg, Am Hubland 97074 W\"urzburg, Germany}
\author{G.~Sangiovanni}
\affiliation{\small\em Institute for Theoretical Physics and Astrophysics, University of W\"urzburg, Am Hubland 97074 W\"urzburg, Germany}
\author{K.~Held}
\affiliation{\small\em Institute for Solid State Physics, TU Wien, 1040 Vienna, Austria}

\date{\small\today}
\begin{abstract}
We derive the improved estimators for general interactions and employ these for  the  continuous-time quantum Monte Carlo method. Using a worm algorithm we show how measuring higher-ordered correlators leads to an improved high-frequency behavior in irreducible quantities such as the one-particle self-energy or the irreducible two-particle vertex for non-density-density interactions. A good knowledge of the asymptotics of the two-particle vertex is essential for calculating non-local electronic correlations using diagrammatic extensions to the dynamical mean field theory as well as for calculating susceptibilities. We test our algorithm against analytic results for the multi-orbital atomic-limit and the Falicov-Kimball model.

\pacs{71.27.+a, 02.70.Ss} 
\end{abstract}
\maketitle

\section{Introduction} \label{sec:Intro}

The Hubbard model~\cite{Hubbard} is one of the most fundamental  models for strong electronic correlations. In the limit of infinite spatial dimensions, an exact mapping onto the Anderson impurity model (AIM)~\cite{Anderson,Hewson} allows for the treatment of local electronic correlations within the framework of dynamical mean field theory (DMFT).~\cite{Metzner,Georges,Kotliar_dmft,Held} For finite spatial dimensions, the aforementioned mapping becomes an approximation; in particular for low-dimensional systems and in the vicinity of second-order phase transitions, non-local correlations beyond DMFT are important.

In order to capture the non-local physics (i.e. \mbox{k-dependent} self-energies, spectral functions etc.) of such lattice models, several extensions to DMFT have been proposed. These extensions can be classified into cluster extensions,~\cite{Hettler} which solve a cluster of sites in a DMFT bath and diagrammatic extensions. While cluster extensions are only capable of capturing non-locality up to the size of the cluster, diagrammatic extension also allow for treating long correlation lengths. Prominent representatives include the dynamical vertex approximation,~\cite{Toschi} the dual fermion approach,~\cite{Rubtsov} the one-particle irreducible approach,~\cite{Rohringer_1PI} and the DMFT to functional renormalization group.~\cite{Taranto}
An extensive treatment of diagrammatic methods and cluster methods for the two dimensional Hubbard model can be found elsewhere.~\cite{LeBlanc}

At the heart of the diagrammatic methods mentioned above lies the local
two-particle vertex as an input which can be calculated from the 
full frequency dependent two-particle susceptibility of the Anderson impurity model. Likewise the calculation of q-dependent susceptibilities in DMFT requires the local vertex or susceptibilities as a starting point. For model calculations and single-orbital systems, the exact diagonalization scheme has proven valuable due to its simplicity, albeit it requires a bath discretization. For more complex systems (i.e. multi-orbital systems and general interactions) at finite temperature, continuous-time quantum Monte Carlo methods~\cite{Rubstov_ct_int,Werner_qmc,Werner,Gull_aux,Gull} are the state-of-the-art impurity solvers. These algorithms stochastically sample the expansion of the imaginary time partition function\cite{Prokofev_ct_orig,Prokofev_ct} and are in principle numerically exact, allowing for general interactions and continuous bath dispersions. When expanding the hybridization in the impurity-bath hybridization (CT-HYB),\cite{Werner_qmc,Werner} this results in the strong-coupling algorithm, which has a favorable scaling over the entire range of interaction strength.\cite{Gull_perfomance}

While the CT-HYB algorithm tends to perform reasonably well regarding the low Matsubara frequency behavior of irreducible quantities such as the self-energy, its high-frequency behavior is usually prone to large statistical fluctuations.\cite{Gull_perfomance} These fluctuations are weaker for the continuous-time algorithm in its interaction expansion (CT-INT)~\cite{Rubstov_ct_int} and in auxiliary fields (CT-AUX).~\cite{Gull_aux} 

Various solutions have been proposed, which can be classified into algorithm-independent methods and others primarily applied to CT-HYB. The former are based on high-frequency expansions of the full and bare Green's functions\cite{Potthoff} resulting in expressions for the asymptotics of the self-energy.~\cite{Gull,Wang} In the context of CT-HYB, proposed methods include noise-filters in the Legendre basis~\cite{Boehnke} which measures higher-ordered correlation functions to yield high precision estimates for the self-energy.~\cite{Hafermann} The latter `improved estimator' technique obtains the self-energy by measuring the two-particle Green's function with three of the four fermionic operators in second quantization at equal (imaginary) times. Combining this quantity with the local interaction yields the self-energy from the equation of motion of the one-particle Green's function. 

Problems in the high-frequency asymptotics are known to exist, not only for the self-energy, but also for the irreducible two-particle vertex. High precision estimates can be obtained by measuring a three-particle Green's function with three of the six fermionic operators at equal times, which so far however, has only been applied for density-density like interactions.~\cite{Hafermann,Hafermann_ret} Here the CT-HYB algorithm further simplifies into its segment representation.~\cite{Werner_qmc} When allowing for non-density-density interactions it becomes much more challenging to calculate higher-ordered correlation functions and one needs to extend CT-HYB by a worm algorithm. 

Previously the worm algorithm was proposed for continuous-time Monte Carlo impurity solvers.~\cite{Burovski,Gunacker} Here essentially both, the partition function and the Green's function, are  expanded in the interaction or the hybridization. The resulting configuration space is enlarged by the different Green's function spaces considered. The concept originates from diagrammatic Monte Carlo solvers for bosonic Green's functions~\cite{Prokofev_ct,Prokofev_hfye}, it was later introduced to the CT-INT algorithm~\cite{Burovski} and the CT-HYB algorithm~\cite{Gunacker}.

In this paper we generalize the worm algorithm in its hybridization expansion to measure the improved estimators for the self-energy and ultimately the connected part of the two-particle Green's function. In Section~\ref{sec:Irr} we introduce our notation and recapitulate the concept of one-particle and two-particle irreducibility and the related Dyson equation or Bethe-Salpeter equation, respectively. Here we also define combined orbital-spin-time indices and channel decompositions. In Section~\ref{sec:Eom} we derive the improved estimators by considering the equation of motion of the one- and two-particle Green's function, employing the path integral formalism~\cite{Wallerberger, Veschgini} instead of the Hamiltonian formalism.~\cite{Hafermann}. In Section~\ref{sec:Worm} we briefly review the concepts of worm sampling in the context of CT-HYB and introduce the Monte Carlo update procedures for the improved estimator worm spaces. In Section~\ref{sec:Results} we compare the one- and two-particle irreducible quantities (self-energy and irreducible two-particle vertex) with the multi-orbital atomic limit for non-density-density interactions. We further consider the Falicov-Kimball (FK) model as a non-trivial system with respect to the CT-HYB algorithm. In particular we also calculate the so-called 'fc'-components of the two-particle Green's function (a propagator describing the interaction between the itinerant and the frozen spin of the FK model, which to the best of our knowledge has not been calculated before). 
Section~\ref{sec:Conclusion} gives a brief summary of the algorithm and results.

\section{Irreducibility and notation}\label{sec:Irr}

Let us first set the stage, briefly introduce our notation and channel decomposition, and summarize the most important relations between the functions considered in our paper.
 On the one-particle level we deal with the interacting Green's $G_{ab}$, the non-interacting one $\mathcal{G}_{ab}$, as well as the self-energy $\Sigma_{ab}$  which are related through the  Dyson equation
\begin{equation}
\label{eq:dyson}
G_{ab} = \mathcal{G}_{ab} + \mathcal{G}_{ac} \Sigma_{cd} G_{db}
\end{equation}
 We use Latin indices from here on to describe a combined index collecting imaginary times $\tau_a$, orbitals $\alpha_a$ (denoted by Greek indices) and spins  ${\sigma_a=\{\uparrow,\downarrow\}}$ into a multi-index $a=(\alpha,\sigma_a,\tau_a)$ or, alternatively, fermionic Matsubara frequencies $\nu_a$ instead of  imaginary times $\tau_a$. We further assume the Einstein summation convention for generalized (Latin) indices, which translates to summation over orbital (Greek) indices and spin indices as well as integration over $\tau\in[0,\beta)$.

At the two-particle level the irreducible vertex function includes all diagrams which are two-particle irreducible, that is, diagrams which cannot be separated by cutting two fermionic lines. When cutting two fermionic 
lines, resulting diagrams can be classified in the particle-particle ($\pp$), the particle-hole ($\ph$) and the transverse particle-hole ($\phbar$) channel,
 following the notation of  Ref.~\onlinecite{Rohringer}. Let us recall that the Bethe-Salpeter 
equation connects the full two-particle vertex  $F$ and  the  irreducible two-particle vertex  $\Gamma^r$ in a channel $r\in\{\pp,\ph,\phbar\}$:
\begin{equation}
\label{eq:bsalpether_general}
F_{abcd} = \Gamma_{abcd}^{r} + \Gamma_{abef}^{r} G_{eg} G_{fh} F_{ghcd},
\end{equation}
where the last term generates the two-particle reducible contributions in a channel $r$.
The above equation couples various spin components of the irreducible vertex $\Gamma^r$. When assuming SU(2) symmetry, one can decouple the above equation by introducing different spin-superpositions.
In this work we consider the density channel (d) and the magnetic channel (m), which are given by (the same holds for $F$)
\begin{align}
\label{eq:channels}
&\Gamma^{\mathrm{d}}_{\alpha \beta \gamma \delta} =  \Gamma_{\alpha \sigma \beta \sigma \gamma \sigma \delta \sigma} + \Gamma_{\alpha \sigma \beta \sigma \gamma (-\sigma) \delta (-\sigma)} \\
&\Gamma^{\mathrm{m}}_{\alpha \beta \gamma \delta} =  \Gamma_{\alpha \sigma \beta \sigma \gamma \sigma \delta \sigma} - \Gamma_{\alpha \sigma \beta \sigma \gamma (-\sigma) \delta (-\sigma)},
\end{align}
where we have omitted the explicit time-dependence. 
Considering SU(2) symmetry the spin-components ${\Gamma_{\sigma (-\sigma) (-\sigma) \sigma} \equiv \Gamma_{\overline{\sigma (-\sigma)}}}$ can be included in the above due to crossing symmetry.
This channel decomposition allows us to write the Bethe-Salpeter equation in a decoupled form (which is very similar to the Dyson equation for the self-energy):
\begin{align}
\label{eq:bsalpether}
\chi^{d,m}_{\alpha \beta \gamma \delta} = \chi^{(0) d,m}_{\alpha \beta \gamma \delta} + \chi^{(0) d,m}_{\alpha \beta \epsilon \zeta} \phantom{.} \Gamma^{d,m}_{\epsilon \zeta \eta \theta} \phantom{.} \chi^{d,m}_{\eta \theta \gamma \delta},
\end{align}
again assuming the time-dependence implicitly. Here,  $\chi$ and $\chi^{(0)}$ denote the susceptibility with and without vertex corrections, respectively. The irreducible vertex $\Gamma^{r}$ resulting from the inversion of the Bethe-Salpeter equation shows
various divergence lines in the metallic phase, which relate to the breakdown of perturbative physics. Divergence lines have been discovered in the Hubbard model~\cite{Schaefer} and the FK model.~\cite{Ribic} Recently a more detailed discussion of
the physical implications of these divergence lines was given.~\cite{SchaeferNonPert} The following calculations were carried out away from any divergence lines.

\begin{figure}
\centering
\includegraphics[scale=0.60]{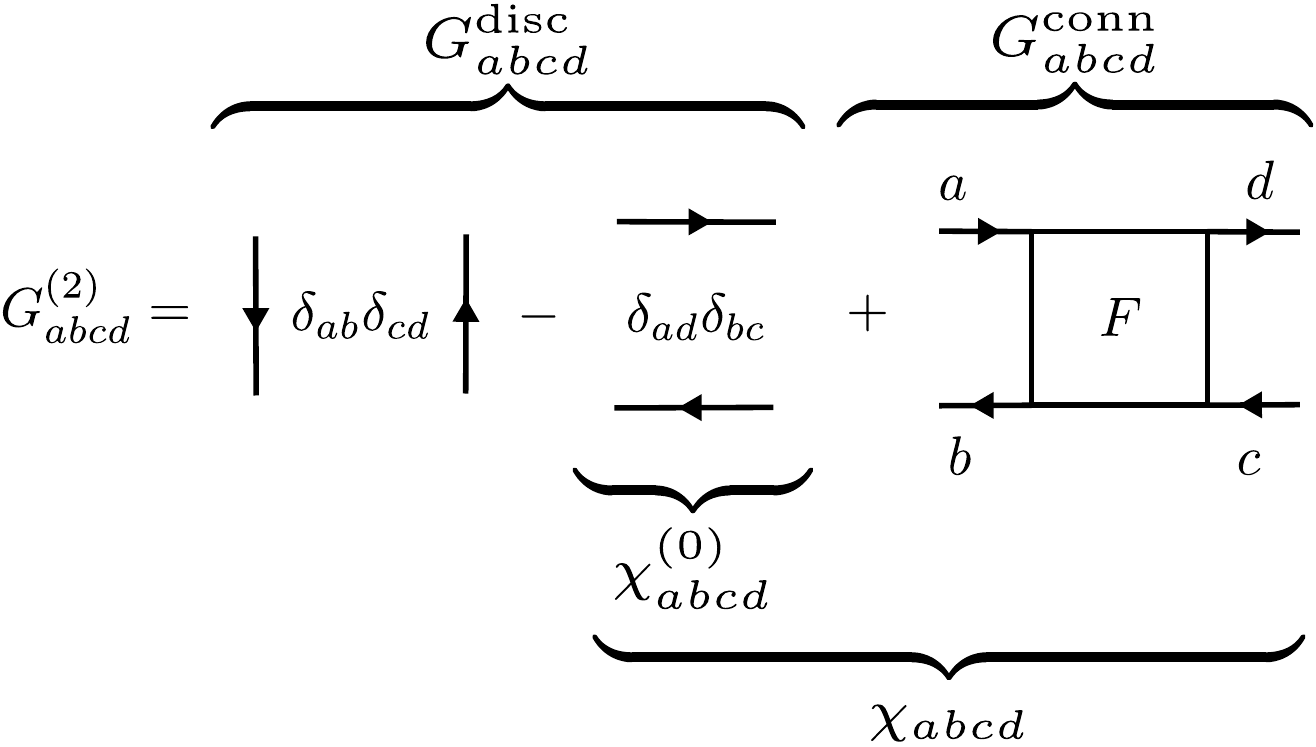}
\caption{Diagrammatic representation of the two-particle Green's function $G^{(2)}$ in terms of the disconnected contribution $G^{\mathrm{disc}}$ and the connected contribution $G^{\mathrm{conn}}$. The alternative partitioning splits the two-particle Green's
function into a generalized susceptibility $\chi$ with a bubble term $\chi^{(0)}$.}
\label{figure:2p_gf} 
\end{figure}

Figure~\ref{figure:2p_gf} illustrates the diagrammatic relation between the two-particle Green's function $G^{(2)}$, the susceptibility and the  two-particle vertex $F$. While usually a definition in terms of a generalized susceptibility $\chi$
with a bubble term $\chi^{(0)}$ is favorable for the Bethe-Salpeter equation~\eqref{eq:bsalpether}, we employ an alternative partitioning into a connected part $G^{\mathrm{conn}}$ and a disconnected part $G^{\mathrm{disc}}$ (see Figure~\ref{figure:2p_gf}), as this will become relevant in the derivation later.

\section{Improved estimators}\label{sec:Eom}

The improved estimators to the self-energy and the irreducible vertex have been
already derived for density-density-type interactions in
Ref.~\onlinecite{Hafermann}. Here, we give a derivation for general interactions which is state-of-the-art for CT-HYB nowadays.  Unlike the original derivation starting from the Hamiltonian, we use
the path integral formalism, loosely following Ref.~\onlinecite{Veschgini} for 
brevity. The partition function of the AIM is  given as
\begin{equation}
\label{eq:z_aim}
Z = \int \mathcal{D}[\bar d,d]\ \ee^{-S [\bar d,d]} ,
\end{equation}
where $\bar{d},d$ are the fermionic Grassmann fields of the impurity electrons.
The action $S$ of the AIM, where the non-interacting bath 
fermions have been integrated out, then reads
\begin{equation}
\label{eq:s_aim}
S =-T[\bar{d},d]+V[\bar{d},d]=-\bar{d}_a \mathcal{G}^{-1}_{ab} d_b + \frac{1}{2} U_{abcd} \bar{d}_a \bar{d}_b d_d d_c,
\end{equation}
where $T[\bar{d},d]$ is the kinetic part and $V[\bar{d},d]$ the interaction part of the action; 
$\mathcal{G}^{-1}_{ab} =- \partial / \partial_{\tau_a} - \epsilon_{ab} - \Delta_{ab}$ is the non-interacting Green's function.
The hybridization function $\Delta_{ab}$, the on-site energies $\epsilon_{ab}$ and the local orbital-dependent interaction $U_{abcd}$ are, in terms of the combined orbital-spin-time index, defined as
\begin{align}
\Delta_{ab} &\coloneqq \Delta_{\alpha \sigma_a \beta \sigma_b}(\tau_a - \tau_b) \\
\epsilon_{ab}      &\coloneqq \epsilon_{\alpha \sigma_a \beta \sigma_b} \delta(\tau_a - \tau_b) \notag \\
U_{abcd}    &\coloneqq U_{\alpha \beta \gamma \delta} \delta_{\sigma_a \sigma_d} \delta_{\sigma_b \sigma_c} \delta(\tau_a - \tau_b) \delta(\tau_a - \tau_c) \delta(\tau_a - \tau_d), \notag
\end{align}
where $\alpha, \beta, \ldots $ are the orbitals of the combined indices $a,b, \ldots \,$.
We remind the reader that the summation convention over repeated (Latin) indices requires the summation over orbital (Greek) indices, spin indices as well as integration over $\tau\in[0,\beta)$.
Using Eqs.~\eqref{eq:z_aim} and~\eqref{eq:s_aim} we can write the one- and two-particle impurity Green's function as
\begin{align}
G^{(1)}_{ab} &= -\frac{1}{Z} \int  \mathcal{D}[\bar{d},d]\ \ee^{-S[\bar{d},d]} d_a \bar{d}_b \label{eq:gf_1}, \\
G^{(2)}_{abcd} &= \frac{1}{Z} \int  \mathcal{D}[\bar{d},d]\ \ee^{-S[\bar{d},d]} d_a \bar{d}_b d_c \bar{d}_d. \label{eq:gf_2}
\end{align}
In order to derive the improved estimators of the self-energy and the vertex function,
we formulate the identity (master equation)
\begin{equation}
 \frac{\mathcal{G}_{ae}}{Z} \int \mathcal{D}[\bar{d},d] \frac{\partial}{\partial \bar{d}_e} e^{-S[\bar{d},d]} F[\bar{d},d] = 0,
\end{equation}
where $F[\bar{d},d]$ is an arbitrary function in $\bar{d}$ and $d$ and $S[\bar{d},d]$ is defined by Eq.~\eqref{eq:s_aim}.
This identity holds true because the integral of the derivative of a Grassmann field vanishes due to the invariance of the path integral
under infinitesimal transformations of this field. A more general discussion of path integrals in a similar framework is found elsewhere.\cite{Zinn} Computing the derivative, we find the Schwinger-Dyson equation in the path integral formalism as~\cite{Veschgini}
\begin{align}
\label{eq:dyson_schwinger}
 &\frac{1}{Z} \int \mathcal{D}[\bar{d},d] \ee^{-S[\bar{d},d]} d_a F[\bar{d},d] = \\
 &\quad\frac{\mathcal{G}_{ae}}{Z} \int  \mathcal{D}[\bar{d},d] \ee^{-S[\bar{d},d]} \left( \frac{\partial V[\bar{d},d]}{\partial \bar{d}_e} F[\bar{d},d]- \frac{\partial F[\bar{d},d]}{\partial \bar{d}_e}\right). \notag 
\end{align}
The derivative of the interaction part in~\eqref{eq:dyson_schwinger} is given by
\begin{equation}
\frac{\partial V[\bar{d},d]}{\partial \bar{d}_e} =  \frac{1}{2} U_{fghi} \left( \delta_{fe} \bar{d}_g  - \bar{d}_{f} \delta_{ge} \right) d_i d_h =: U_{[eg]hi} \bar{d}_f d_i d_h,
\end{equation}
where the square brackets $[...]$ denotes the antisymmetrization over the indices (including a factor $\frac{1}{2}$). 

By choosing $F[\bar{d},d]$ properly we can generate improved estimators up to an arbitrary order of Green's functions. The important cases of the self-energy and two-particle vertex function are discussed in the next two section.

\subsection{Self-energy}

\begin{figure}
\centering
\includegraphics[scale=0.60]{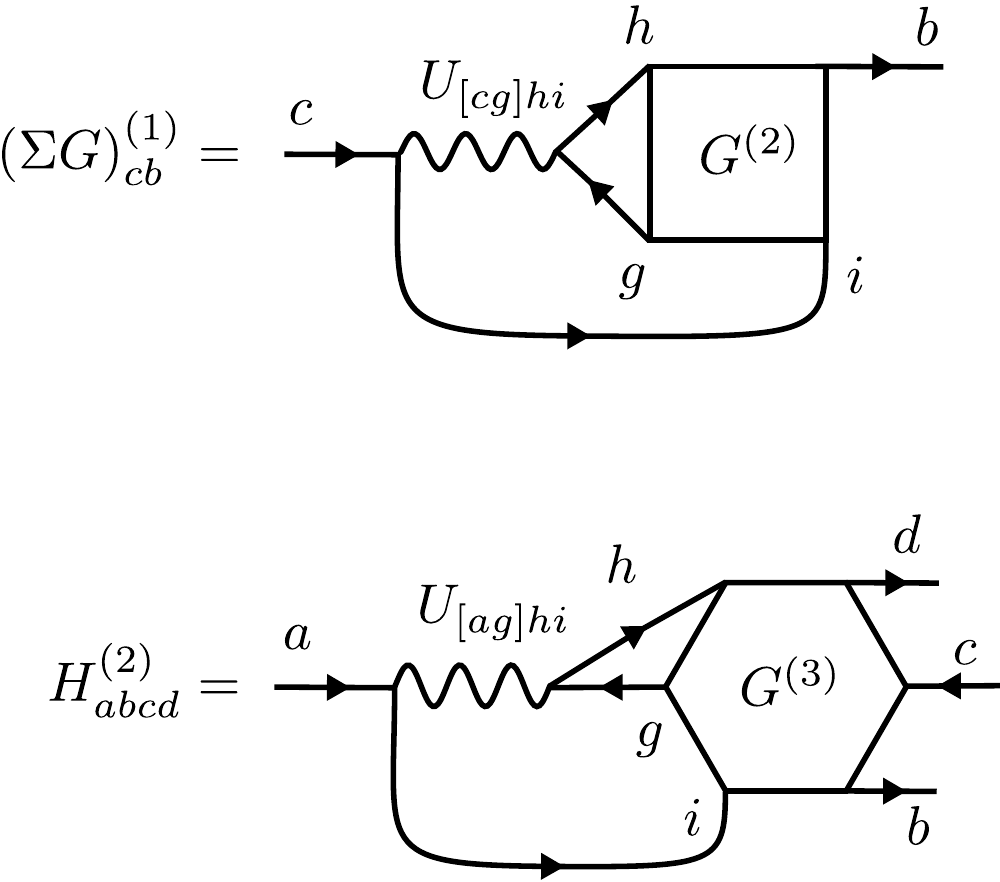}
\caption{Top: diagrammatic representation of the one-particle improved estimator $(\Sigma G)^{(1)}_{bc}$ [Eq.~\eqref{eq:sigmag_path}].
Bottom: diagrammatic representation of the two-particle improved estimator $H^{(2)}_{abcd}$ [Eq.~\eqref{eq:h_expectation}, the last part of Eq.~\eqref{eq:h_path}]. The local
interaction is represented explicitly by a wiggly line.}
\label{figure:estimator_diagramms} 
\end{figure}

In order to obtain an estimator for the self-energy we set  $F[\bar{d},d]=\bar{d}_b$ in Eq.~\eqref{eq:dyson_schwinger}, recovering
the one-particle Green's function~\eqref{eq:gf_1} on the left hand side 
and the following  right hand side
\begin{equation}
 G_{ab} = \mathcal{G}_{ab} - \frac{\mathcal{G}_{ac}}{Z} \int \mathcal{D}[\bar{d},d] \ee^{-S[\bar{d},d]} U_{[cg]hi} \bar{d}_g d_i d_h \bar{d}_b.
\end{equation}
Comparing this with the Dyson equation~\eqref{eq:dyson} we find
\begin{equation}
\label{eq:sigmag_path}
 \left(\Sigma G \right)_{cb} = -\frac{1}{Z} \int \mathcal{D}[\bar{d},d] \ee^{-S[\bar{d},d]} U_{[cg]hi} \bar{d}_g d_i d_h \bar{d}_b.
\end{equation}
The diagrammatic representation of this one-particle improved estimator is given in Figure~\ref{figure:estimator_diagramms} (top).  Let us now recall the explicit indices from the combined Latin indices and rewrite the path integral in second quantization as a thermal expectation value
\begin{multline}
\label{eq:sigmag_expectation}
 \left(\Sigma G \right)_{\alpha \sigma, \beta \sigma^\prime}\!(\tau-\tau^\prime) = \\
 - \langle  T_\tau\, U_{[\alpha \gamma] \delta \epsilon}
 \,\ddgr_{\gamma\sigma^{\dprime}}\!(\tau) \,\dee_{\epsilon\sigma}\!(\tau)
 \,\dee_{\delta\sigma^{\dprime}}\!(\tau) \,\ddgr_{\beta\sigma^\prime}\!(\tau^\prime) \rangle,
\end{multline}
where we have introduced the time-ordering symbol $T_\tau$ and switched from fermionic Grassmann variables $\bar{d},d$ to creation and annihilation operators $\ddgr, \dee$. 
In making the imaginary time index explicit, we find that the spontaneous nature of the interaction contracts three operators to a single (imaginary) time.  In terms of computational complexity the calculation of the one-particle improved estimator is thus comparable to the one-particle Green's function.

\subsection{Vertex function}

In order to obtain an estimator for the vertex-function we set $F[\bar{d},d]=-\bar{d}_b d_c \bar{d}_d$ in Eq.~\eqref{eq:dyson_schwinger}, so that the left hand
side becomes the two-particle Green's function~\eqref{eq:gf_2}:
\begin{align}
  G_{abcd} &= \mathcal{G}_{ab} G_{cd} - \mathcal{G}_{ad} G_{bc} + \\
 &\frac{\mathcal{G}_{ae}}{Z} \int \mathcal{D}[\bar{d},d] \ee^{-S[\bar{d},d]} U_{[eg]hi} \bar{d}_g d_i d_h \bar{d}_b d_c \bar{d}_d \notag.
\end{align}
We multiply the above with $\mathcal{G}^{-1}_{aj}$ from the left and apply the Dyson equation $\mathcal{G}^{-1}_{aj} = G^{-1}_{aj} + \Sigma_{aj}$
\begin{align}
  \left( G^{-1}_{aj} + \Sigma_{aj} \right) & G_{abcd} = \delta_{jb} G_{cd} - \delta_{jd} G_{bc} + \\
 &\frac{\delta_{je}}{Z} \int \mathcal{D}[\bar{d},d] \ee^{-S[\bar{d},d]} U_{[eg]hi} \bar{d}_g d_i d_h \bar{d}_b d_c \bar{d}_d \notag.
\end{align}
In the following we multiply with $G_{ja}$ from the left and finally rearrange the terms
\begin{align}
\label{eq:h_path}
&G_{abcd} - G_{ab} G_{cd} + G_{ad} G_{bc}  =  -(G \Sigma)_{ae} G_{ebcd} + \\
& + \frac{G_{ae}}{Z} \int \mathcal{D}[\bar{d},d] \ee^{-S[\bar{d},d]} U_{[eg]hi} \bar{d}_g d_i d_h \bar{d}_b d_c \bar{d}_d. \notag
\end{align}
We can identify the left-hand side with the connected part $G^{\mathrm{conn}}$ of the two-particle Green's function, see Figure~\ref{figure:2p_gf}.
The diagrammatic representation of the two-particle improved estimator is given in Figure~\ref{figure:estimator_diagramms} (bottom).
We observe that we are required to obtain the one-particle estimator $(G\Sigma)$ apart from sampling the two-particle improved estimator. The final result yields
\begin{align}
\label{eq:g_conn}
&G^{\mathrm{conn}}_{abcd}  =  -(G \Sigma)_{ae} G_{ebcd} + G_{ae} H_{ebcd}.
\end{align}
For the two-particle improved estimator we recover the explicit indices from the combined Latin indices and rewrite the remaining path integral of Eq.~\eqref{eq:h_path} as a thermal expectation value in second quantization
\begin{align}
\label{eq:h_expectation}
& H_{\alpha \sigma_a, \beta \sigma_b, \gamma \sigma_c, \delta \sigma_d }(\tau_a,\tau_b,\tau_c,\tau_d) = \langle T_\tau\,U_{[\alpha \epsilon] \zeta \eta}  \times \\ 
& \quad \ddgr_{\epsilon\sigma_e}\!(\tau_a) \,\dee_{\eta\sigma_a}\!(\tau_a) 
      \,\dee_{\zeta\sigma_e}\!(\tau_a) \,\ddgr_{\beta\sigma_b}\!(\tau_b)
      \,\dee_{\gamma\sigma_c}\!(\tau_c) \,\ddgr_{\delta\sigma_d}\!(\tau_d)
      \,\rangle. \notag
\end{align}
Again, by making the imaginary time index explicit, we find that three operators are contracted to a single time, whereas the other three operator have each a different time argument.
In terms of computational complexity the two-particle improved estimator is hence comparable to the two-particle Green's function.

\section{Worm sampling}\label{sec:Worm}

The expectation values in Eqs.~\eqref{eq:sigmag_expectation} and~\eqref{eq:h_expectation} are already in the correct form
required by the worm sampling algorithm of CT-HYB. We will further assume a diagonal hybridization function $\Delta_{\alpha\sigma,\beta\sigma^\prime} = \Delta_{\alpha\sigma,\alpha\sigma} \delta_{\alpha\beta}\delta_{\sigma\sigma^\prime}$ in
order to allow for a well-behaved sign in the CT-HYB algorithm.
For diagonal hybridization, all one-particle quantities have a single spin-orbit degree of freedom, i.e.  $G_{\alpha\sigma,\beta\sigma^\prime} = G_{\alpha\sigma,\alpha\sigma} \delta_{\alpha\beta}\delta_{\sigma\sigma^\prime}$. Consequently, $(\Sigma G) = (G\Sigma)$.

\begin{figure}
\centering
\includegraphics[scale=0.34]{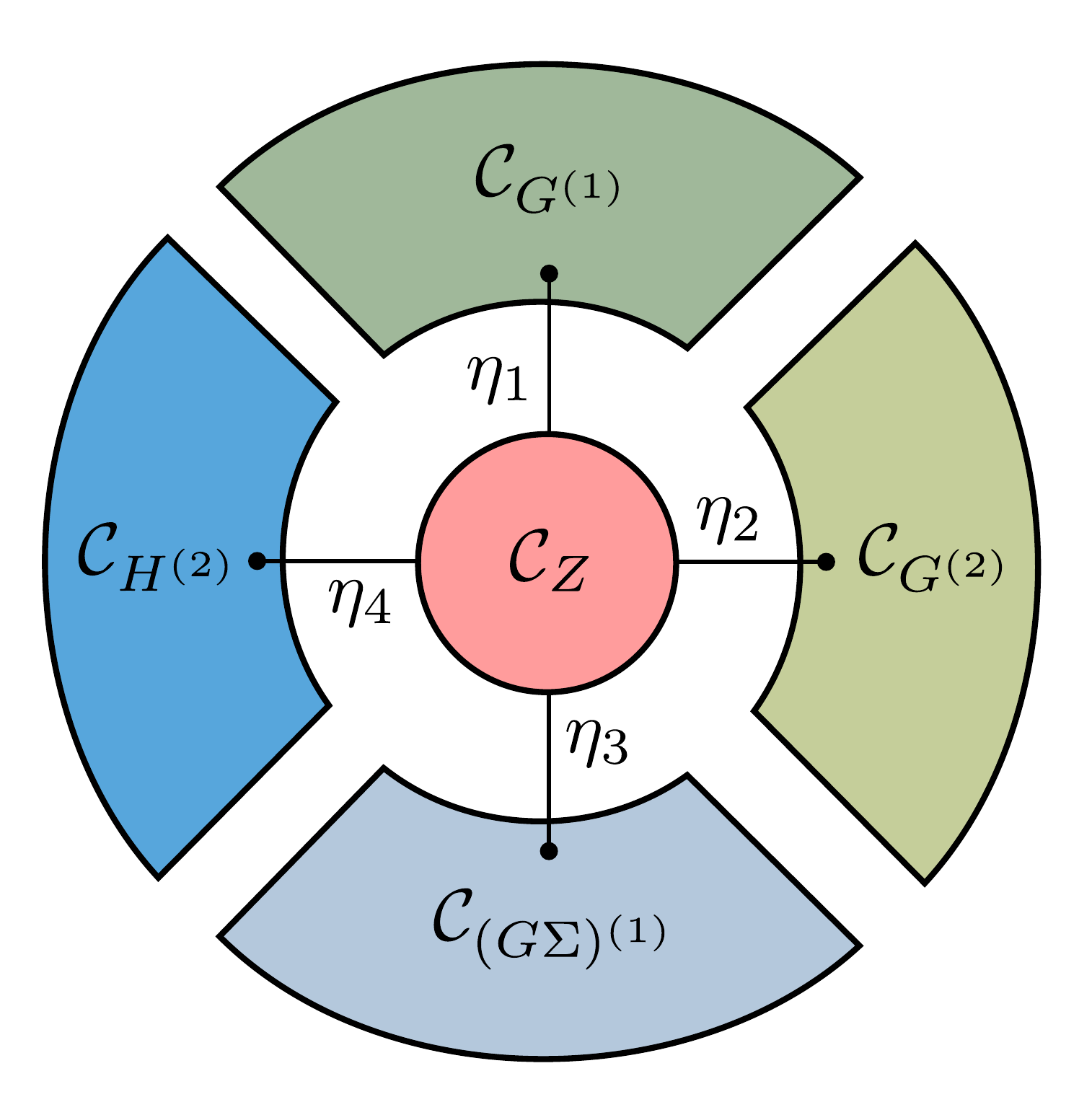}
\caption{Schematic representation of the extended configuration space including the partition function space $\mathcal{C}_Z$ (red), the Green's function spaces $\mathcal{C}_{G^{(1)}}$ and $\mathcal{C}_{G^{(2)}}$ (green) and the improved estimator spaces $\mathcal{C}_{(G\Sigma)^{(1)}}$ and $\mathcal{C}_{H^{(2)}}$ (blue). Worm spaces are only linked to
one another over the partition function space with the balancing parameters $\eta_i$.}
\label{figure:configspace} 
\end{figure}

The basic idea of worm sampling is to extend the configuration space to include the worm spaces of interest. With respect to $G^{\mathrm{conn}}$
in Eq.~\eqref{eq:g_conn} this results in an enlarged configuration space
\begin{equation}
\label{eq:config_ext}
 \mathcal{C} = \mathcal{C}_Z \oplus \mathcal{C}_{G^{(1)}} \oplus \mathcal{C}_{G^{(2)}} \oplus \mathcal{C}_{(G\Sigma)^{(1)}} \oplus \mathcal{C}_{H^{(2)}},
\end{equation}
where $\mathcal{C}_Z$ is the partition function space extended by the worm spaces as illustrated in Figure~\ref{figure:configspace}. For more details and an introduction to worm sampling, see Ref.~\onlinecite{Gunacker}.
To distinguish the worm spaces further, we will refer to $\mathcal{C}_{G^{(1)}}$ and $\mathcal{C}_{G^{(2)}}$ as Green's function spaces and
to $\mathcal{C}_{(G\Sigma)^{(1)}}$ and $\mathcal{C}_{H^{(2)}}$ as improved estimator spaces.

In order to jump between partition function space and worm spaces we introduce worm insertion and removal operators. While in principle jumping directly between
worm spaces is possible, we only allow for worm spaces to be connected via the partition function space $\mathcal{C}_Z$. 
More specifically, a direct connection between the Green's function space $\mathcal{C}_{G^{(2)}}$ and the improved estimator 
space $\mathcal{C}_{(G\Sigma)^{(1)}}$ can be established by shift moves which set the three worm operators connected by the interaction $U$ to an equal time.
However, we will not follow this route, as not all components present in $\mathcal{C}_{G^{(2)}}$ are necessarily present in $\mathcal{C}_{(G\Sigma)^{(1)}}$
due to the interaction term $U_{\alpha \beta \gamma \delta}$.
Directly connecting $\mathcal{C}_{G^{(1)}}$ and $\mathcal{C}_{G^{(2)}}$ results in severe ergodicity problems due to quantum number rejects.
For Slater-Kanamori like interactions, this approach does not recover spin-flip and pair-hopping terms as already discussed in Ref.~\onlinecite{Gunacker}.

We introduce the partition function of the extended configuration space as
\begin{multline}
\label{eq:part_ext}
 W = Z + \eta_1 Z_{G^{(1)}} + \eta_2 Z_{G^{(2)}}  + \eta_3 Z_{(G\Sigma)^{(1)}} + \eta_4 Z_{H^{(2)}}.
\end{multline}
Here, the balancing parameters $\eta_i$ to sample each fraction of the extended configuration space with equal likelihood.

\subsection{Reducing the extended configuration space}

Looking at the generalized partition function $W$~\eqref{eq:part_ext} and the extended configuration space $\mathcal{C}$~\eqref{eq:config_ext} we
observe that keeping track of the different worm spaces becomes quite involved.
Following Ref.~\onlinecite{Hafermann}, we can define the one-particle Green's function $G^{(1)}$ with respect to the one-particle
improved estimator $G\Sigma^{(1)}$ and the non-interacting Green's function $\mathcal{G}$ by employing the Dyson equation
\begin{equation}
\label{eq:g_from_ie}
G^{(1)}_{ab} =  \mathcal{G}_{ac} \underbrace{\left(\delta_{cb} + (\Sigma G)^{(1)}_{cb} \right)}_{\equiv  A^{(1)}_{cb}}.
\end{equation}
For the connected part resulting from the two-particle improved estimator in~\eqref{eq:g_conn} we substitute $G_{abcd} = G^{\mathrm{disc}}_{abcd} +G^{\mathrm{conn}}_{abcd}$ and find
\begin{align}
\label{eq:g_conn_red}
G^{\mathrm{conn}}_{ebcd}  &=  (\delta_{ae} + (G\Sigma)_{ae}^{(1)})^{-1} \left(-(G \Sigma)_{ae} G^{\mathrm{disc}}_{ebcd} + G_{ae} H_{ebcd} \right) \notag\\
 &= (A_{ae}^{(1)\dagger})^{-1} \left(-(G \Sigma)_{ae} G^{\mathrm{disc}}_{ebcd} + G_{ae} H_{ebcd} \right).
\end{align}
where $G^{\mathrm{disc}}$ can be calculated using the one-particle Green's function  defined in~\eqref{eq:g_from_ie}. With the above rewriting we effectively reduce
the worm spaces from four to two. That is, we only sample the improved estimator spaces $\mathcal{C}_{(G\Sigma)^{(1)}}$ and $\mathcal{C}_{H^{(2)}}$, and do not need to consider the Green's function spaces explicitly.

\subsection{Worm insertion and removal steps in improved estimator space}

The Metropolis acceptance rate of the one-particle improved estimator introduced in~\eqref{eq:sigmag_expectation} is given by
\begin{multline}
\label{eq:metropolis_1p}
 a(\mathcal{C}_Z \rightarrow \mathcal{C}_{(G\Sigma)^{(1)}_{ab}}) = \\
 \mathrm{min}\!\left[1, \eta_3  \frac{\left| U_{[\alpha \gamma] \delta \epsilon} w_{\mathrm{loc}}(n+4,\tau_{i_1},\ldots,\tau_{i_n};\tau_a, \tau_b) \right|}
{\left| w_{\mathrm{loc}}(n,\tau_{i_1},\ldots,\tau_{i_n}) \right|} 
\beta^{4} \right]\!.
\end{multline}
where in this work $n$ is the number of operators in the local trace at times $\tau_{i_1} \ldots \tau_{i_n}$ and $w_{\mathrm{loc}}$ is the local part
of the configuration weight. 
In principle it is possible to move the interaction term $U_{\alpha \gamma \delta \epsilon}$ of the Metropolis acceptance rate into the Monte Carlo estimator during the measurement.
In this case, however, one needs to sample different components of the implicit summation over the three equal time operators explicitly, while otherwise it is possible to define a new operator, which
sums up all components beforehand.

Similar to the Metropolis acceptance rate of the one-particle improved estimator, the acceptance rate of the two-particle
improved estimator introduced in~\eqref{eq:h_expectation} is given by
\begin{multline}
\label{eq:metropolis_2p}
 a(\mathcal{C}_Z \rightarrow \mathcal{C}_{H^{(2)_{abcd}}}) = \mathrm{min}\!\left[1, \eta_4 \times \right.\\
 \left.  \frac{\left| U_{[\alpha \epsilon] \zeta \eta} w_{\mathrm{loc}}(n+6,\tau_{i_1},\ldots,\tau_{i_n};\tau_a, \tau_b, \tau_c, \tau_d) \right|}
{\left| w_{\mathrm{loc}}(n,\tau_{i_1},\ldots,\tau_{i_n}) \right|} 
 \beta^{6} \right]\!.
\end{multline}
We emphasize that the underlying idea of worm sampling in CT-HYB is to continue the sampling of operators connected to hybridization lines inside the worm spaces. Similar as the series expansion of the partition
function with respect to the hybridization, one may think of a hybridization expansion of the observable in question, but now with the additional external worm operators, that are not connected by hybridization lines. In the worm algorithm this series is
then sampled stochastically just as one would sample the partition function.
Merely inserting and removing worm operators without further sampling results in a non-ergodic sampling procedure, as some diagrams cannot be generated in this way.\cite{Gunacker}

\subsection{Worm measurement}

The measurement of observables in worm spaces is trivially determined by recording imaginary time bins during the Monte Carlo sampling ($\langle \ldots \rangle_{\mathrm{MC}}$) for a given spin-orbital component and only needs
to be corrected in its normalization and sign (sgn), see Ref.~\onlinecite{Gunacker} for further technical details:
\begin{equation}
\label{eq:binning}
(G\Sigma)^{(1)}_{\mathcal{C}_{(G\Sigma)}}(\tau-\tau') = -\langle \mathrm{sgn}(U w_{\mathrm{loc}}) \phantom{.} \delta(\tau-\tau')\rangle_\mathrm{MC}.
\end{equation}
or equivalently in Matsubara frequencies:
\begin{equation}
\label{eq:unbinned}
(G\Sigma)^{(1)}_{\mathcal{C}_{(G\Sigma)}}(i\nu) = \langle \mathrm{sgn}(U w_{\mathrm{loc}}) \phantom{.} \ee^{i\nu(\tau-\tau')}\rangle_\mathrm{MC}.
\end{equation}
Similarly, the two-particle improved estimator in the particle-hole convention is measured as
\begin{multline}
\label{eq:unbinned_2p}
H^{(2)}_{\mathcal{C}_H}(i\nu, i\nu', i\omega) = \\
\qquad\langle \mathrm{sgn}(U w_{\mathrm{loc}}) \phantom{.} \ee^{i\nu(\tau_1-\tau_2)} \ee^{i\nu'(\tau_3-\tau_4)} \ee^{i\omega(\tau_2-\tau_3)}\rangle_\mathrm{MC}.
\end{multline}
It is important to note that the sign of the configuration now includes an additional sign from the interaction term $U_{\alpha \beta \gamma \delta}$,
which was introduced to the Metropolis acceptance rate in~\eqref{eq:metropolis_1p} and~\eqref{eq:metropolis_2p}.
We point out that the sign problem of the worm algorithm is identical to the sign problem of the hybridization expansion itself. 
That is, the average sign in the denominator of the estimators originates from the normalization with respect to the partition function, i.e. being a consequence of the average sign of partition function space.

While Eq.~\eqref{eq:binning} may be binned in imaginary time $\tau$, and afterwards Fourier transformed to Matsubara frequencies $i\nu$, the unbinned Fourier transform
in Eq.~\eqref{eq:unbinned} is possible as well.
In case of the two-particle quantities a binning procedure becomes much more involved as one needs to generate a grid which further resolves the sign changes due to anticommutating operators.
Thus, employing a nonequispaced fast Fourier transform algorithm\cite{Nfft} in Eq.~\eqref{eq:unbinned_2p} is preferable.

The correct normalization of observables measured in any of the worm spaces is given by
\begin{equation}
 \langle \mathcal{A} \rangle = \frac{1}{\eta_i} \frac{N_{\mathcal{C}_\mathcal{A}}}{N_Z} \langle \mathcal{A} \rangle_{\mathcal{C}_{\mathcal{A}}},
\end{equation}
where $\langle \mathcal{A} \rangle$ is the expectation value of the operators $G^{(1)},G^{(2)},(G\Sigma)^{(1)}$ or $H^{(2)}$ with physical normalization
and $\langle \mathcal{A} \rangle_{\mathcal{C}_{\mathcal{A}}}$ is the corresponding expectation value with its worm space normalization. Further, $N_{\mathcal{C}_A}$ is the number of steps taken in the configuration space $\mathcal{C}_A$ and $N_Z$
is the number of steps taken in partition function space $\mathcal{C}_Z$.

\section{Results}\label{sec:Results}

\subsection{Atomic limit}\label{sec:Atomic}

The atomic limit is defined for an arbitrary lattice, where the hopping of electrons between different sites vanishes. This is equivalent to the AIM where the hybridization function $\Delta_{ij}$ vanishes
for all spin-orbit components. Up to this point we have not specified the local interaction. We will investigate an SU(2)-symmetric Slater-Kanamori interaction given by~\cite{Kanamori} 
\begin{align}
\label{eq:slater_kanamori}
H_{\mathrm{loc}} &= \sum_{\alpha} U \hat n_{\alpha\uparrow} \hat n_{\alpha\downarrow} \\
&+ \sum_{\alpha > \beta, \sigma} \left[ U' \hat n_{\alpha\sigma} \hat n_{\beta(-\sigma)} 
             + (U' - J) \hat n_{\alpha\sigma} \hat n_{\beta\sigma}\right] \notag \\
&- \sum_{\alpha \neq \beta} J \left(\ddgr_{\alpha\downarrow} \ddgr_{\beta\uparrow} \dee_{\beta\downarrow} \dee_{\alpha\uparrow} 
   + \ddgr_{\beta\uparrow} \ddgr_{\beta\downarrow} \dee_{\alpha\uparrow} \dee_{\alpha\downarrow} + \mathrm{h.c.} \right), \notag
\end{align}
where ${\hat n_{\alpha\sigma} \coloneqq \ddgr_{\alpha\sigma} \dee_{\alpha\sigma}}$  denotes the density operator. 
We have made the sums explicit here in order to represent $U_{\alpha \beta \gamma \delta}$ by the inter-orbital repulsion $U$, the intra-orbital repulsion $U'$ and the interaction due to Hund's coupling by $J$.

The Slater-Kanamori interaction contains spin-flip and pair-hopping terms, which translate to further non-vanishing components in the two-particle Green's function and improved estimators. 
These components are not accessible through the removal of hybridization lines, as in the case of the traditional CT-HYB approach. 
Worm sampling allows us to calculate \textit{all} correlation functions independent of the details of the interaction $U_{\alpha \beta \gamma \delta}$ and the hybridization $\Delta_{ab}$.
As a result, the above algorithm is especially suited for material calculations with less symmetries in the interaction (e.g. the full Coulomb interaction).
Due to the computational effort involved when calculating two-particle quantities, however, multi-orbital calculations usually assume SU(2)-symmetric interactions and employ the PS quantum number~\cite{Parragh} and diagonal hybridization functions in order to avoid any sign problems.

Here we consider the half-filled two-orbital atomic limit. The interaction parameters are set to $U\equiv 1.0$ (setting our unit of energy), $U'=0.5$ and $J=0.25$. The half-filling condition for the Slater-Kanamori interaction is given by
$\mu = \frac{3}{2} U - \frac{5}{2} J$, such that $\mu= 0.875$. The inverse temperature was set to $\beta = 10$.

\begin{figure}
\centering
\fontsize{12}{12}\selectfont
\resizebox{1.0\columnwidth}{!}{\input{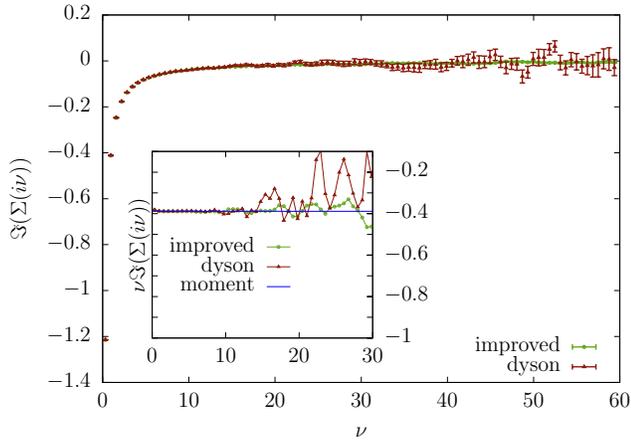}}
\caption{Imaginary part of the self-energy $\Sigma$  vs.\ Matsubara frequencies $i\nu$ in the atomic limit with two half-filled orbitals, $U=1.0$, $U'=0.5$ and $J=0.25$. The self-energy obtained from the Dyson equation (red) shows much larger
fluctuations in the high-frequency region compared to the self-energy obtained from the improved estimators (green). Errorbars are calculated from 40 bootstrap samples. Inset: Comparing the second moment of the self-energy from the improved estimator (green) and the Dyson equation (red) by multiplication with the Matsubara frequency with the measurement via the
one- and two-particle density matrix(blue).\cite{Wang}}
\label{figure:atomic_siw} 
\end{figure}

Figure~\ref{figure:atomic_siw} shows the imaginary part of the self-energy in fermionic Matsubara frequencies.
Both the improved estimator $(\Sigma G)^{(1)}(i\nu)$ and the Green's function $G^{(1)}(i\nu)$ were obtained using worm sampling with comparable computational effort.
We observe  large fluctuations in the high-frequency region when calculating the self-energy from the Dyson equation.
Calculating the self-energy from the improved estimator instead, yields  a much better high-frequency behavior in 
Figure~\ref{figure:atomic_siw}.

\begin{figure}
\centering
\fontsize{12}{12}\selectfont
\resizebox{1.0\columnwidth}{!}{\input{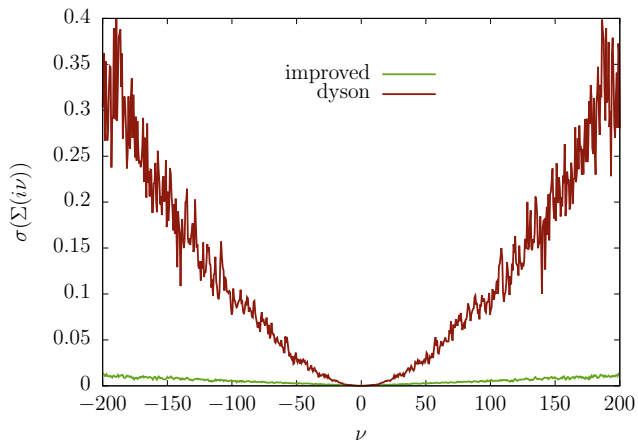}}
\caption{Error of the self-energy $\Sigma$ vs.\ Matsubara frequencies $i\nu$ in the atomic limit for the data given in Figure~\ref{figure:atomic_siw}. The uncertainties resulting from the Dyson equation (red) follow a quadratic scaling, while the uncertainties obtained
from the improved estimator (green) follow a weak linear scaling.}
\label{figure:err} 
\end{figure}

The contrasting high-frequency behavior is a consequence of a different propagation of statistical uncertainties.
Empirically, we find roughly constant errorbars for both the one-particle Green's function $G^{(1)}(i\nu)$ and the one-particle improved estimator $(\Sigma G)^{(1)}(i\nu)$ over the entire frequency range.
Performing a formal error propagation for the self-energy through the Dyson equation, we find the statistical fluctuations of the self-energy 
diverge quadratically for large frequencies. This is consistent with the red curve in Figure~\ref{figure:err}. If we instead consider the error propagation for the self-energy assuming instead the improved estimator $\Sigma G$ simply being
multiplied by $G^{-1}$ from the right, we find a linear scaling of the statistical uncertainties over the frequency range.

\begin{figure}
\centering
\fontsize{12}{12}\selectfont
\resizebox{1.0\columnwidth}{!}{\input{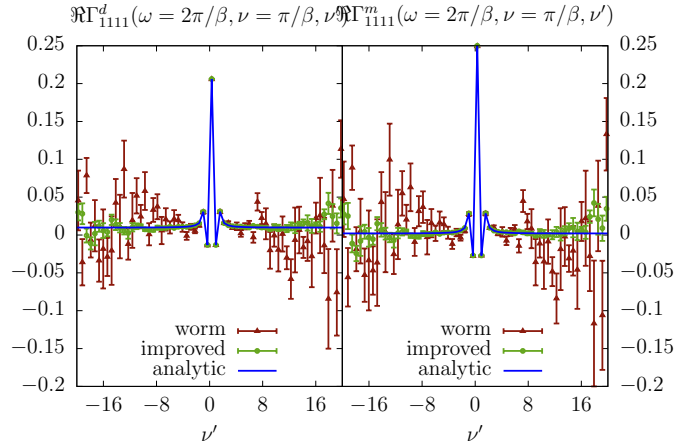}}
\caption{Fermionic cut through the  two-particle vertex function in the density (left) and magnetic (right) channel for the same parameters as in Figure~\ref{figure:atomic_siw}. The vertex obtained from a straight-forward calculation of the 
two-particle Green's function (red) displays larger fluctuations in the high-frequency region than the result obtained from the two-particle improved estimator (green). Errorbars are calculated from 4 independent inversions. The analytic result (blue) is shown for comparison.}
\label{figure:atomic_irreducible_1} 
\end{figure}

Figure~\ref{figure:atomic_irreducible_1} shows the irreducible two-particle vertex function in the density and magnetic channel, in comparison with the exact solution.~\cite{Rohringer,Hafermann_atomic} To this end the connected Green function~\eqref{eq:g_conn} was calculated from this the susceptibility and finally the irreducible vertex through the Bethe-Salpeter equation~\eqref{eq:bsalpether}.
A much better high-frequency behavior in the two-particle improved estimator here allows for a more stable inversion of the Bethe-Salpeter equation.
This better high-frequency  behavior is obtained by using the improved estimator
which hence dramatically reduces the error of the vertex in Figure~\ref{figure:atomic_irreducible_1}.

At the one-particle level we could explain the high-frequency fluctuations by replacing the Dyson equation with the Schwinger-Dyson equation in a formal error propagation.
At the two-particle level there is no corresponding substitution for the Bethe-Salpeter equation. Hence, a similar formal argument is not available. Nonetheless we observe that also in this case the improved estimator reduces the error considerably in Figure~\ref{figure:atomic_irreducible_1}.

\subsection{Falicov-Kimball (FK) model}\label{sec:FK}

The FK model~\cite{Falikov} can be seen as a non-trivial extension to the atomic limit, where one spin is frozen (vanishing hopping), while the
other spin is itinerant (non-vanishing hopping). The Hamiltonian of the single-orbital spin-less FK model reads
\begin{multline}
\label{eq:hamiltonian_fk}
H_{\mathrm{FK}} = -t \sum_{ij} \hat{c}_i^\dagger \hat{c}_j + U \sum_i \hat{c}_i^\dagger \hat{c}_i \hat{f}_i^\dagger \hat{f}_i \\ 
                 - \mu \sum_i \hat{c}_i^\dagger \hat{c}_i - \epsilon_f \sum_i \hat{f}_i^\dagger \hat{f}_i,
\end{multline}

where $t$ denotes the hopping amplitude from site $j$ to $i$ of itinerant $c$-electrons, $U$ the local Coulomb repulsion between an itinerant $c$-electron and a frozen $f$-electron. Further,
$\mu$ and $\epsilon_f$ are local potentials of the itinerant and localized electrons.
In the context of DMFT, the FK model maps onto the self-consistent solution of the resonant level model (RLM),~\cite{Brandt} which is Eq.~\eqref{eq:hamiltonian_fk} with $U$ and $\epsilon_f$ restricted to a single site.
Aside from an analytic expression for the $c$-electron self-energy, in general the propagators (Green's functions) of the $c$-electrons
are also analytically accessible. Propagators involving the $f$-electrons on the other hand are much more difficult to obtain. More detailed information about the FK model
can be found in Ref.~\onlinecite{Freericks}.

In terms of the CT-HYB algorithm, the FK model is specifically challenging, because the traditional formulation of the algorithm
is not capable of directly measuring the propagators for any $f$-electrons due to the vanishing hybridization function.
The worm algorithm allows for sampling and measuring the $f$-electrons, and is thus the natural formulation of a FK solver in the context of CT-HYB.

In the following we investigate a two-dimensional FK model out of half-filling with inverse temperature $\beta=20$, interaction parameter $U=1.0$ and chemical potential $\mu=0.2$, where the half bandwidth $D\equiv 1$ of the conduction electrons sets our unit of energy.
In order to fix the $f$-occupation to $n_f=0.25$ (in terms of the RLM $p_1=0.25$) we adjust the $f$-electron energy level to $\epsilon_f=-0.038114$. 

\begin{figure}
\centering
\fontsize{12}{12}\selectfont
\resizebox{1.0\columnwidth}{!}{\input{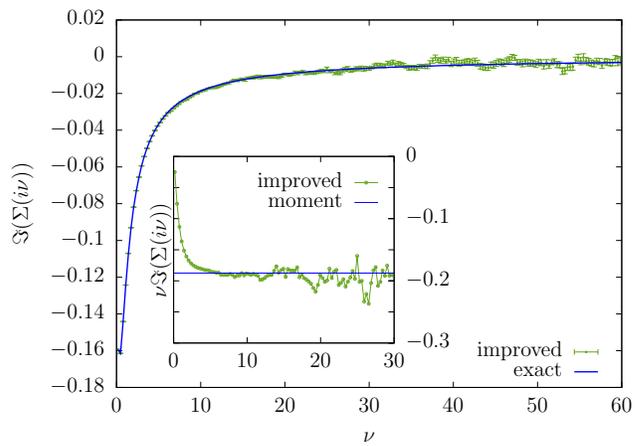}}
\caption{Imaginary part of the self-energy $\Sigma$ vs.\ Matsubara frequencies $i\nu$ for the FK model at $U=1.0$,  $\beta=20$, $\mu=0.2$, and  $\epsilon_f=-0.038114$ ($n_f=0.25$). 
The self-energy obtained from the improved estimators (green) is in good agreement with the exact self-energy obtained from the RLM (blue). Errorbars are calculated from 40 bootstrap samples. Inset: Comparing the second moment of the self-energy from the improved estimator (green) by multiplication with the Matsubara frequency with the analytical calculation (blue).}
\label{figure:falikov_siw} 
\end{figure}

Figure~\ref{figure:falikov_siw} shows the self-energy of the $c$-electrons for the FK model obtained from DMFT and the improved estimators. The CT-HYB data have fluctuations in the high-frequency
region, but these are well-behaved. This in principle allows us to combine the low-frequency region of the improved estimators with the asymptotic high energy behavior which can be obtained through analytic 
equations from the density. The latter in turn can be calculated during the same run.

When comparing the connected part of the two-particle Green's function for the $c$-electrons $G^{\mathrm{conn}}_{(cc)}$ we find a good agreement
of our CT-HYB improved estimator with the exact result~\cite{Ribic} (not shown).

\begin{figure}
\centering
\includegraphics[scale=0.10]{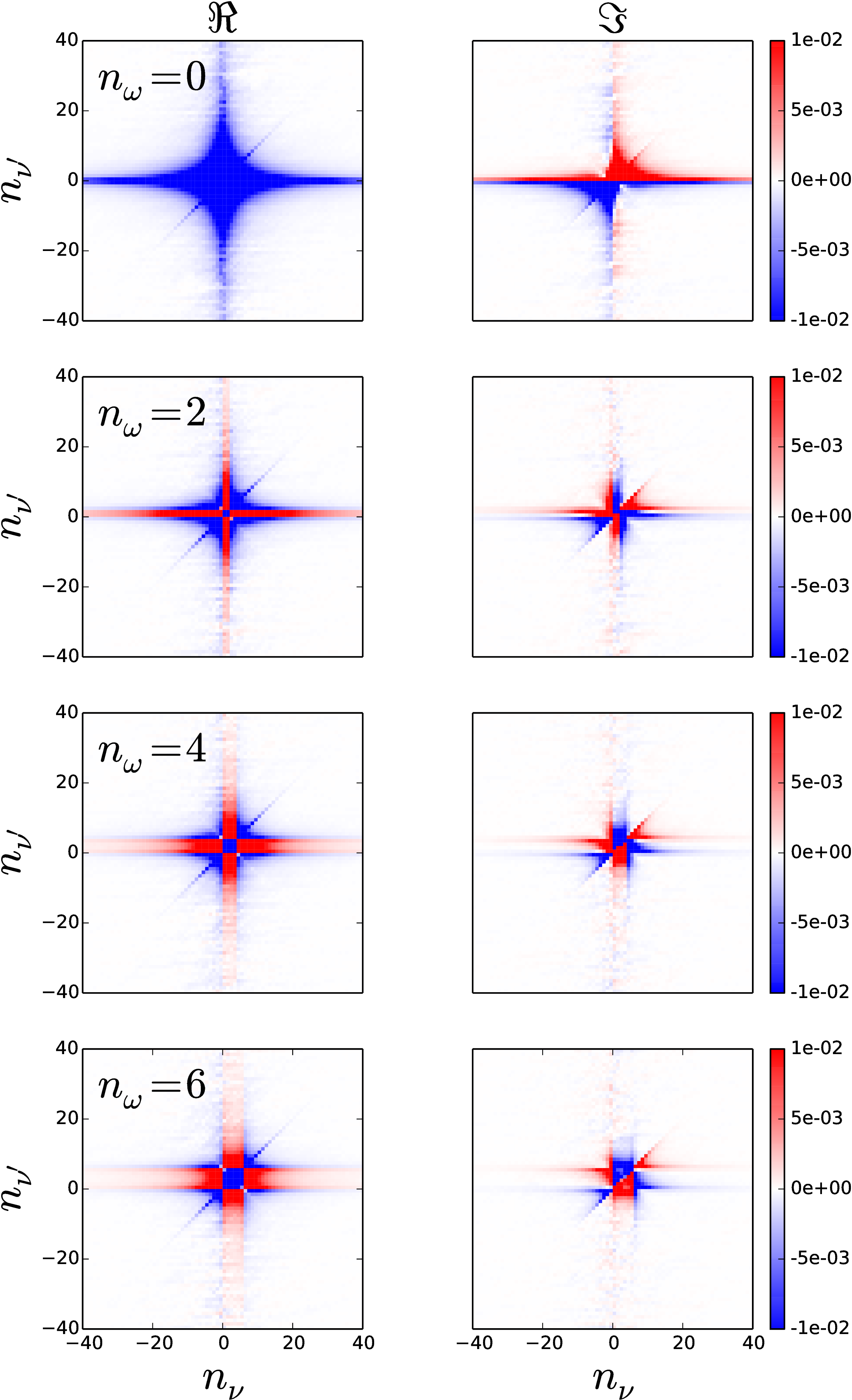}
\caption{Real and imaginary part of the connected part of the Green function, $G^{\mathrm{conn}}_{(fc)}$, for different bosonic frequencies $n_\omega = 0,2,4,6$ and the same parameters as in Figure~\ref{figure:falikov_siw}. All Matsubara frequencies are given in terms of their (integer) index.}
\label{figure:falikov_conn} 
\end{figure}

Figure~\ref{figure:falikov_conn} shows our CT-HYB results for the connected part $G^{\mathrm{conn}}_{(fc)}$ of the FK model, which cannot be obtained analytically 
in a straight-forward way.  We observe typical ``cross'' and ``plus'' structures
in the real and imaginary part. Sign changes in the connected part can be observed. These structures shift and broaden with increasing bosonic frequency. Outside these structures the connected part vanishes.

The validity of the $fc$-component of the two-particle Green's function is indicated, albeit implicitly, since the calculation of the equal-time (or equivalently frequency-summed) 
component enters the equation of motion for the self-energy of the $c$-electrons, which we found to agree with the analytical result in Figure~\ref{figure:falikov_siw}. A correct self-energy hence implies that the $fc$-component of the two-particle Green's function
is equally correct.
Please note that calculating the $fc$-component of the irreducible two-particle vertex is more involved as the FK model violates the SU(2)-symmetry. Thus, a channel decomposition is no longer possible and the Bethe-Salpeter equation~\eqref{eq:bsalpether_general} does not decouple anymore, mixing $fc$- and $ff$-components of the irreducible vertices.

\section{Conclusion}\label{sec:Conclusion}

In this work we have presented a generalization of the improved estimator scheme for the CT-HYB algorithm.
We make use of the recently introduced worm-algorithm in CT-HYB to sample the necessary equal-time correlators.
This allows us  to treat general interactions beyond the density-density type.
We demonstrate that the improved estimator scheme has, compared to the direct calculation, a superior convergence in the high-frequency region for the self-energy and the irreducible
two-particle vertex function. The atomic limit for a two-orbital model with non-density-density interaction is used to validate our numerical CT-HYB results against analytical expressions. We demonstrate the necessity of the worm algorithm to numerically calculate all propagators of the FK model. 
Specifically,  results for the density and magnetic channel of the irreducible two-particle vertex are supplied, which can be used as an input for diagrammatic methods beyond DMFT as well as to calculate q-dependent susceptibilities within DMFT.
We strongly emphasize that the improved estimators formulated in terms of the worm algorithm will greatly enhance multi-orbital material calculations employing non-density-density interactions.

\acknowledgments
We thank P. Thunstr\"om for valuable discussions.
This work has been supported  by the  Vienna Scientific Cluster (VSC) Research Center funded by the Austrian Federal Ministry of Science, Research and Economy (bmwfw), the Deutsche Forschungs Gemeinschaft (DFG) through research unit FOR 1346, and the European Research Council under the European Union's Seventh Framework Programme (FP/2007-2013)/ERC grant agreement n. 306447 (AbinitioD$\Gamma$A).  
A. H. and G.S. have been supported by the DFG (through SFB 1170 ``ToCoTronics'').
 The computational results presented have been achieved using the VSC.

\bibliography{bibliography}
 
\vfill\eject

\end{document}